\documentclass[11pt,preprint]{aastex}

\usepackage{graphicx}
\usepackage{float}
\usepackage{color}
\usepackage{verbatim}
\usepackage{lscape}

\shorttitle{Cosmic Radio Background}
\shortauthors{Subrahmanyan \& Cowsik}

\begin{document}


\title{Is there an unaccounted excess Extragalactic Cosmic Radio Background?}


\author{Ravi Subrahmanyan\altaffilmark{1} }
\affil{Raman Research Institute, CV Raman Avenue, Sadashivanagar, Bangalore 560080, India}
\email{rsubrahm@rri.res.in}

\and

\author{Ramanath Cowsik}
\affil{{Physics Department and McDonnell Center for the Space Sciences, Washington University, \\Campus Box 1105, St. Louis, MO 63130}}
\email{cowsik@physics.wustl.edu}


\altaffiltext{1}{Visiting Scientist, National Radio Astronomy Observatory, P O Box 0, Socorro, NM 87801, USA.  The National Radio Astronomy Observatory is a facility of the National Science Foundation operated under cooperative agreement by Associated Universities, Inc.}


\begin{abstract}
Analyses of measurements of the distribution of absolute brightness temperature over the radio sky have led recently to suggestions that there exists a substantial unexplained extragalactic radio background.  Consequently, there have been numerous attempts to place constraints on plausible origins for the `excess'.  We suggest here this expectation of a large extragalactic background, over and above that contributed by the sources observed in the surveys, is based on an extremely simple geometry adopted to model the Galactic emission and the procedure adopted in the estimation of the extragalactic contribution. In this paper, we derive the extragalactic radio background from wide-field radio images by using a more realistic modeling of the Galactic emission and decompose the sky maps at 150, 408, and 1420 MHz into anisotropic Galactic and isotropic extragalactic components. The anisotropic Galactic component is assumed to arise from a highly flattened spheroid representing the thick disc, embedded in a spherical halo, both centered at the Galactic center, along with Galactic sources, filamentary structures, Galactic loops and spurs. All components are constrained to be positive and the optimization scheme minimizes the sky area occupied by the complex filaments. We show that in contrast to the simple modeling of the Galactic emission as a plane parallel slab,  the more realistic modeling yields estimates for the uniform extragalactic brightness that is consistent with expectations from known extragalactic radio source populations.  
\end{abstract}


\keywords{cosmic background radiation - diffuse radiation - radio continuum: general - radio continuum: ISM - 
methods: data analysis}



\section{Introduction}

The cosmic radio background \citep{Jan33,Cla70} represents the sum total of diffuse and discrete sources of emission in the Milky Way galaxy, in external galaxies and in the intergalactic medium as well as any cosmological radiation backgrounds that are redshifted into the radio window.  Since its discovery, the distribution of the brightness temperature of the radio background has been imaged over substantial areas of the sky and in multiple frequencies (see \citet{deO08} for a recent compilation).  Additionally, deep surveys have explored the discrete radio source populations down to $\mu$Jy flux densities and deduced their contribution to the radio background \citep{Ger08, Ver11,Con12}.

The Absolute Radiometer for Cosmology, Astrophysics, and Diffuse Emission 2 (ARCADE 2) measurements of the absolute sky brightness at frequencies 3, 8 and 10~GHz, as well as previous measurements at lower frequencies, were analyzed by \citet{Kog11} using a plane parallel slab model for the Galactic emission. They concluded that the Galaxy contributed only about 30\% to the radio intensities observed at high latitudes. After removing this estimate for the foreground emission from our Galaxy and the cosmic microwave background (CMB) from the total emission, \citet{Fix11} and also \citet{Sei11} find that the residual exceeds the integrated contribution of the known population of extragalactic radio sources by factors of five or more.

With a goal of understanding the origin for this unexplained `excess' cosmic radio background, \citet{Sin10} examined several plausible diffuse and discrete extragalactic source populations that were not contained in surveys to date and concluded that the bulk of the high latitude sky surface brightness might arise from faint and high-redshifted populations at flux density levels hitherto unexplored.  A similar conclusion was arrived at by \citet{Ver11} based on detailed modeling of plausible  extrapolations to measurements of radio source counts at multiple frequencies.  Motivated by these conclusions, \citet{Hol12} examined the observational constraints on clustering of distant radio source populations based on limits on arcmin-scale anisotropy in the radio background and argued against an origin for the excess in objects associated with collapsed  dark matter halos at moderate redshifts.  More recently,  a deep survey with the Karl G. Jansky Very Large Array (VLA) excluded an origin associated with normal galaxies at cosmological distances \citep{Con12}.  The idea that an unexplained `excess' cosmic radio background exists has motivated dark matter interpretations for its origin \citep{For11} and has also been cited as a potential consequence of gravitational wave model for dark energy \citep{Bie13}.

\section{Plane parallel slab model for the Galactic emission}

The assumption that the Galactic emission may be modeled as arising from a plane parallel disc has been critical to the conclusion drawn by \citet{Kog11} that there exists a substantial extragalactic radio flux of unknown origins. Such a model will yield a run of sky brightness temperature $T_A$ versus Galactic latitude $b$, having the linear form $T_A(|b|)=a_0+a_1\csc(|b|)$, that was fit to the distribution of pixel temperatures of ARCADE 2 data, as well as the maps at lower frequencies, to obtain the slope $a_1$ and the intercept $a_0$. 
The slope $a_{1}$  is interpreted to be the value of the brightness temperature of the slab component toward the Galactic pole.  The intercept $a_{0}$ is the mean brightness of the remainder.  

A plane parallel slab model was also adopted in the analysis of the all-sky maps of differential temperature made by the Wilkinson Microwave Anisotropy Probe (WMAP).  The maps were first processed to remove the CMB anisotropy including the dipole, after which the pixel brightness in their all-sky maps at each frequency were fit with such a slab model and scaled to set the intercept $a_{0}$ to zero \citep{Ben03}. In these fits the image pixels at low Galactic latitudes were omitted since the model presumably may not be appropriate along lines of sight close to the plane.  Additionally, WMAP analysis also masked image pixels with substantial contamination from foreground structures.  

Examples of such fits of plane parallel slab models are, for example, in Fig.~3 of \citet{Kog11} and Fig.~7 of \citet{Ben03}.  The goodness of such fits indicates that the mean or median brightness of the sky does indeed decrease linearly with csc($ | b | $), at least to first order.  However, it may be noted here that the Galactic emission does have substantial structure in the form of loops and spurs, which are evident in any low frequency image of the sky as, for example, the 408-MHz all sky map of \citet{Has82}.   The fit of a slab model may be understood to be averaging over these structures for the reason that the average brightness of the sky binned by csc($ | b | $) averages over these structures and the interstitial sky areas of relatively lower brightness.   The fit hence does not represent the totality of the Galactic emission but only an average emission.  A subtraction of such a slab model described by a single parameter that represents the brightness toward the Galactic Pole would consequently leave a remainder that would have substantial contamination by Galactic structures.  This remainder has an average brightness that is given by the fit intercept $a_{0}$, which is not to be interpreted as extragalactic brightness because of this contamination; in fact, there would inevitably be an `excess' in this intercept value when compared to expectations for the uniform extragalactic brightness.

\begin{figure}
\centering
\includegraphics[width=9.0cm]{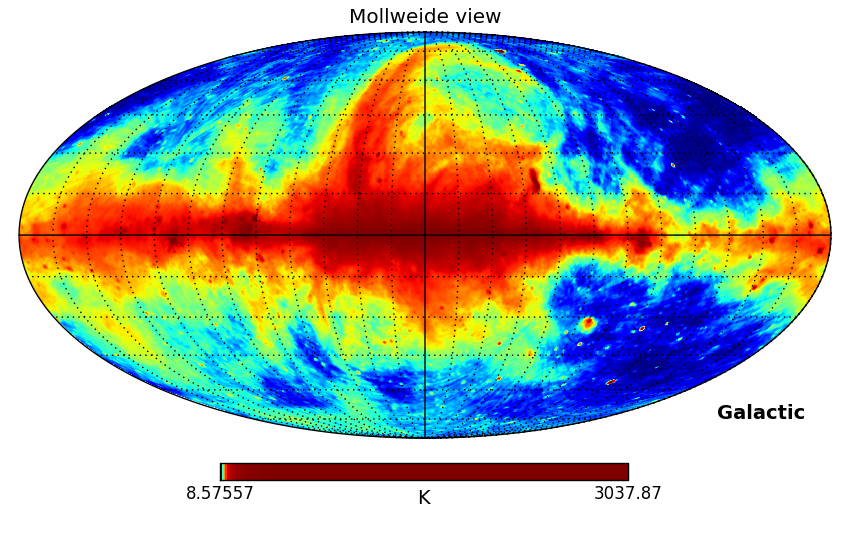}
\includegraphics[width=9.0cm]{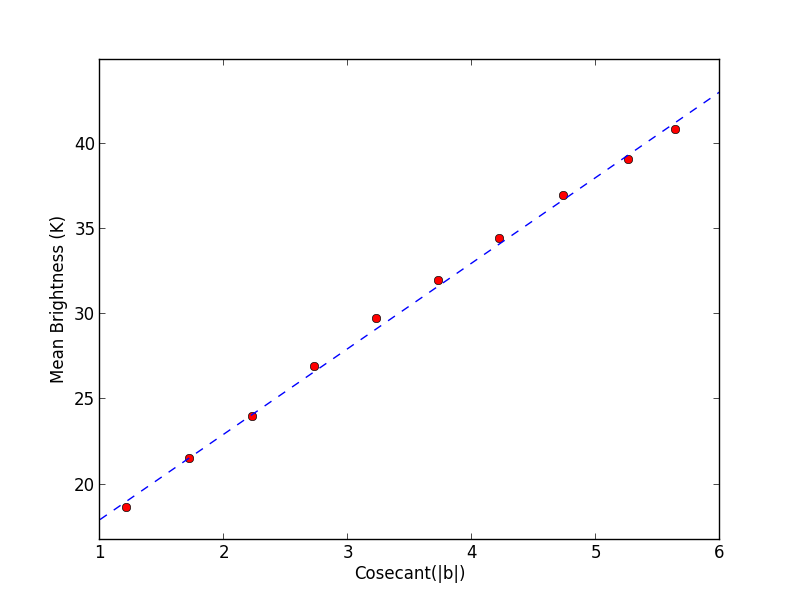}
\includegraphics[width=9.0cm]{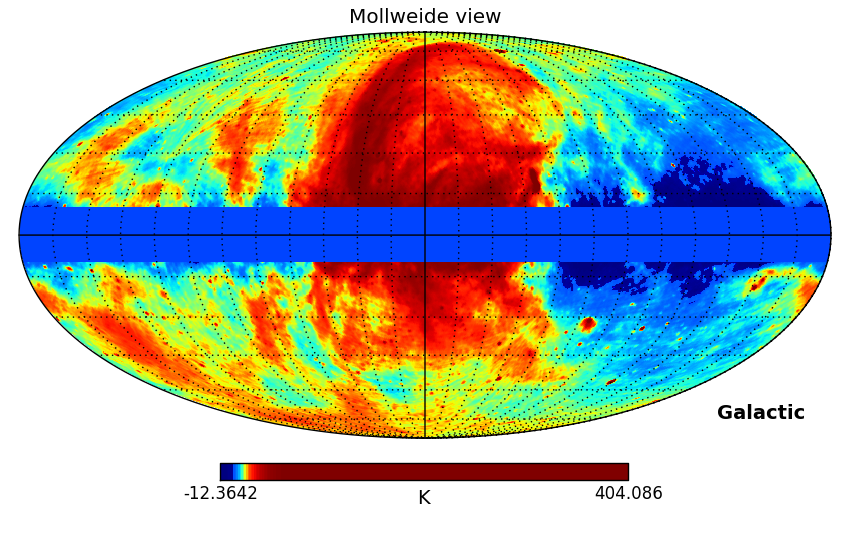}
\caption{ A demonstration of the inadequacy of the procedure adopted previously. The 408-MHz all-sky map is shown in the top panel.  The mean brightness binned in  csc($ | b | $) is shown in the middle panel along with a straight line fit to the data.  The residual image following subtraction of a plane parallel slab model with brightness towards the pole corresponding to the slope of the fit is shown in the bottom panel. (The images are histogram equalized and in the bottom panel pixels with latitude $|b| < 10\degr$ is arbitrarily set to zero.) The average of the bottom panel clearly cannot be taken to represent the extragalactic contribution. }     
\label{fig:slab_model}
\end{figure}

We show in the top panel of Fig.~\ref{fig:slab_model} the 408-MHz all-sky map made by \citet{Has82}.  The CMB monopole corresponding to a brightness temperature of 2.725~K has been subtracted throughout. The image pixels at latitudes $|b| > 10\degr$ were binned and the mean brightness is plotted against csc($ | b | $) in the middle panel of the figure.  The straight line fit to the data yields a slope of 5.0~K and the intercept is 12.9~K; the slope is consistent with the 408-MHz brightness derived by \citet{Kog11} for the Galactic component towards the pole assuming such a slab model.  The residual image following subtraction of the slab model for the Galactic component is shown in the bottom panel.  The region within $\pm 10\degr$ latitude is set to zero brightness for the display.  There are regions of the residual image that have negative brightness and substantial residual Galactic structure remains.  The mean over this residual image at latitudes $|b| > 10\degr$ is estimated to be $\sim12.9$~K, which agreed with the intercept. This intercept is just the average over the residuals in the all sky map and, by no means represents the extragalactic brightness.  

\section{Towards an effective modeling procedure}

\begin{figure}
\centering
\includegraphics[width=12cm]{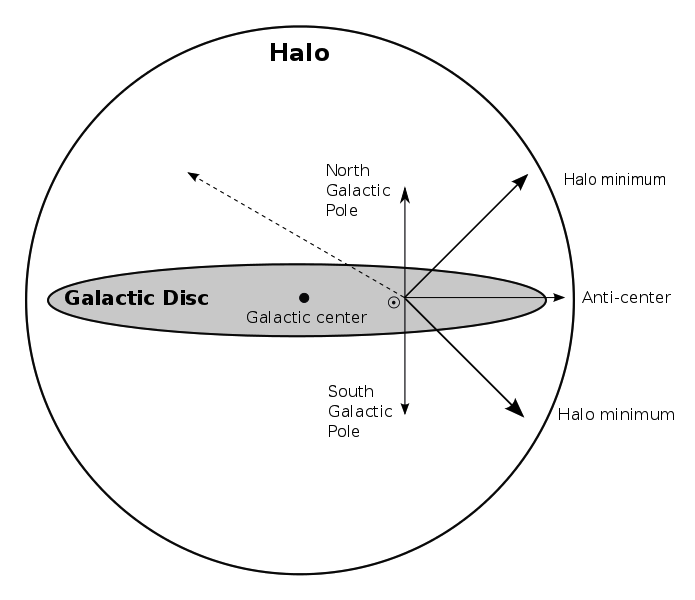}
\caption{Sketch of the model adopted in the present analysis showing regions generally designated as the Galactic disc and halo, even though there exists no sharp boundaries. From the location of the Sun, the observed emission will be anisotropic, including that from the nearby spherical halo, except for a reflection symmetry about the Galactic plane. Directions of the Galactic center, anti-center, North and South Galactic Poles and the halo minima are also indicated.} 
\label{fig:galaxy_model}
\end{figure}

Accordingly, the estimation of the extragalactic component requires a procedure that includes, \emph{ab initio}, an isotropic component apart from other Galactic components in fitting the all-sky maps. This is the crux of the new procedure we have adopted here. The model for the Galactic emission should also be improved in order to get the best estimates of the isotropic extragalactic components.
 \citet{Beu85} presented a two-disc model for the Galactic radio emission and inferred that 90\% of the total power at 408~MHz is in a thick disc, which extends beyond 1.5 times the radius of the solar circle and has an axial ratio of about 5:1.  A plane parallel disc model or a highly flattened spheroidal disc with a high axial ratio would have a minimum close to the Galactic poles. However, it has been known for several decades that the average brightness of the radio sky has a minimum that is offset from the poles toward the direction of the Galactic anti-center, which suggests that any model for the Galactic radio emission requires a halo that extends beyond the solar circle.  \citet{Web75} presented evidence that modeling of the gross radio emission distribution over the sky requires a uniform extragalactic brightness plus a disc and halo centered at the Galactic center.  We may refer to \citet{Sal88} for a comprehensive review of Galactic nonthermal continuum emission.

Consistent with these previous works that have bases in observational data, we have adopted a four component model for the radio sky (see Fig.~\ref{fig:galaxy_model}).  The extragalactic component is expected to be of uniform brightness over the entire sky.  
The Galaxy with its disc and halo components, which are axisymmetric about an axis passing through the Galactic center yield anisotropic intensity distributions because the Sun is located in the Galactic plane about 8.3 kpc away from the center. Motivated by a wide range of astronomical observations such as the mass distribution of the Galaxy, the distribution of supernova remnants, etc., the disc is modeled as a highly-flattened spheroid (i.e., any meridional section through the center is an ellipse of high eccentricity) and a spherical halo, both centered at the Galactic center. To this we add a fourth component, which comprises all the remainder, namely, a complex sky distribution that represents the contributions of the Galactic sources, loops, spurs and filamentary emission.  The flattened spheroid is modeled by  the radial extent in the plane and the axial ratio, and the sphere is modeled by its radius; all these dimensions are in units of the radius of the solar circle, $\omega_\odot$.  The disc and the halo are each assumed to have uniform emissivity, which may be different.  In all, six parameters describe the decomposition at any frequency: 1) semi-major axis of the spheroid, 2) its minor axis, 3) the radius of the halo, 4) the emissivity of the disc, 5) the emissivity of the halo, and 6) the intensity of the isotropic extragalactic radio background.

In the work presented here we have separately modeled the sky maps at each frequency.  Joint modeling of multi-frequency images is reserved for future work.

\section{A multi-frequency decomposition of the cosmic radio background}

The data for the modeling is all-sky maps of the brightness temperature distribution at 150, 408 and 1420 MHz.  The 150 MHz map is from \citet{Lan70} and the 408 MHz map is from \citet{Has82}.   The 1420 MHz map is a combination of the northern sky surveys by \citet{Rei82} and \citet{Rei86} and the southern sky survey by \citet{Rei01}.  We have chosen not to use Fourier filtered versions of the 408~MHz map (available in the LAMBDA\footnote{http://lambda.gsfc.nasa.gov/} website) and the de-striped 1420~MHz map made by \citet{Pla03} because although the filtering is expected to mitigate scanning stripes in the image, any Fourier filtering is equivalent to smoothing in the image plane and may potentially bias the background level, which is the focus of the present work.  The 150 MHz image has a resolution of $2\fdg2$ where as the 408 MHz and 1420 MHz maps were made using beam full width at half maximum (FWHM) of $0\fdg8$ and $0\fdg6$ respectively.  The higher frequency maps were convolved to a final resolution of $1\fdg0$ and all three images were set in HEALPIX format \citep{Gor05} with R8 pixel coordinates (this corresponds to a pixel spacing of $0\fdg229$).  The CMB monopole corresponding to a uniform brightness temperature of 2.725~K was subtracted from all three images prior to fitting with the models.

The model fitting followed by a decomposition of the all-sky maps was carried out for each frequency separately.  Key to the decomposition is the choice of the statistic that is optimized or, in this case, minimized.  We wish to resolve the cosmic radio background into the aforementioned four components---spheroidal disc and spherical halo, extragalactic background and a remainder---with the constraints that the remainder, which represents the finer aspects of the Galactic structure, is positive within errors and that it occupies minimal sky area.  These two criteria together compel the three regular components to maximally fit the data without exceeding the total sky emission.  Separately, a threshold $\tau$ is defined, which represents the acceptable error in the modeling; this threshold is progressively reduced in successive iterations that refine the model parameters and converge on a final parameter set.  For any choice of parameters that define the uniform background and the disc and the halo components, an image corresponding to the remainder is computed.  The statistic to be minimized is chosen to be the number of positive pixels exceeding the threshold $\tau$, plus ten times the number of negative pixels below $-\tau$ .  The factor ten strongly excludes negative pixels and minimization of this statistic corresponds to minimization of the sky area occupied by the remainder while driving the model to best fit the data within $\pm \tau$.

The first step in the decomposition is a global optimization of the model parameters using simulated annealing \citep{Kir83}.   In units of the radius of the solar circle $\omega_\odot\approx8.3$~kpc, the semi-major axis of the ellipsoid in the plane was {\it a priori} limited to the range 1.5--2.5 and the radius of the spheroidal halo was {\it a priori} limited to the range 1.5--3.0.  The ratio of semi-major to semi-minor axes of the ellipsoid was limited to the range 3--6.  At 1420 MHz, the emissivity of the ellipsoid and halo were limited, respectively, to 0.05--2.0 and 0.025--1.0 K per path length of solar circle radius; additionally, the constant extragalactic brightness was allowed to be in the range 0.01--1.0 K.   The emissivity and brightness priors at lower frequencies were obtained by scaling these limits using a temperature power-law index of $-2.6$.   The global optimization using annealing used images degraded to R5 resolution, corresponding to pixel spacing of $1\fdg832$, to speed up the computation.

The solution from annealing was used as an initial guess for a subsequent optimization of the parameters using the downhill simplex algorithm \citep{Nel65}, which was implemented with the threshold for the optimization statistic reduced by a factor of two.  This was followed by another run of simplex optimization that used the image with R7 resolution, corresponding to pixel spacing of $0\fdg458$, and with the threshold reduced by another factor of two.  The final threshold $\tau$ was chosen to be 0.025~K at 1420~MHz; the threshold was also scaled using a temperature index of $-2.6$ to set thresholds at lower frequencies.

\begin{landscape}
\begin{figure}
\begin{minipage}[b]{0.33\linewidth}
\includegraphics[width=\linewidth]{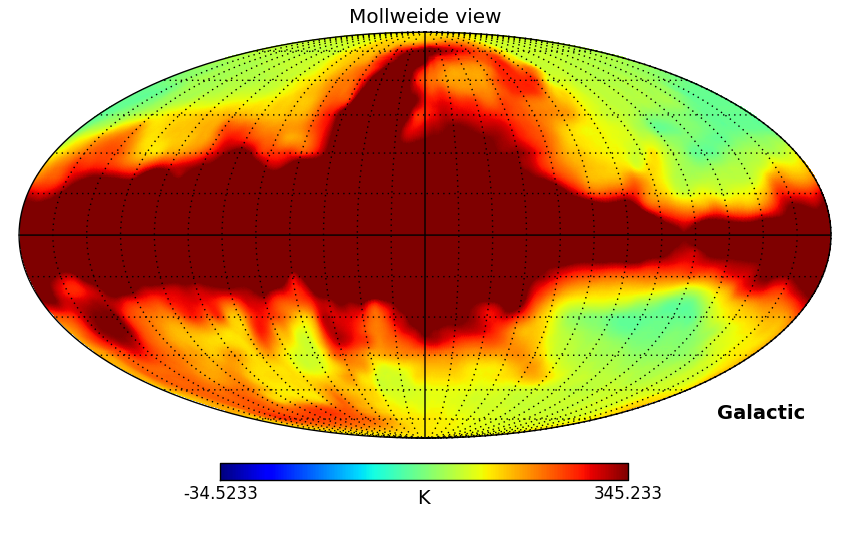}
\end{minipage}
\begin{minipage}[b]{0.33\linewidth}
\includegraphics[width=\linewidth]{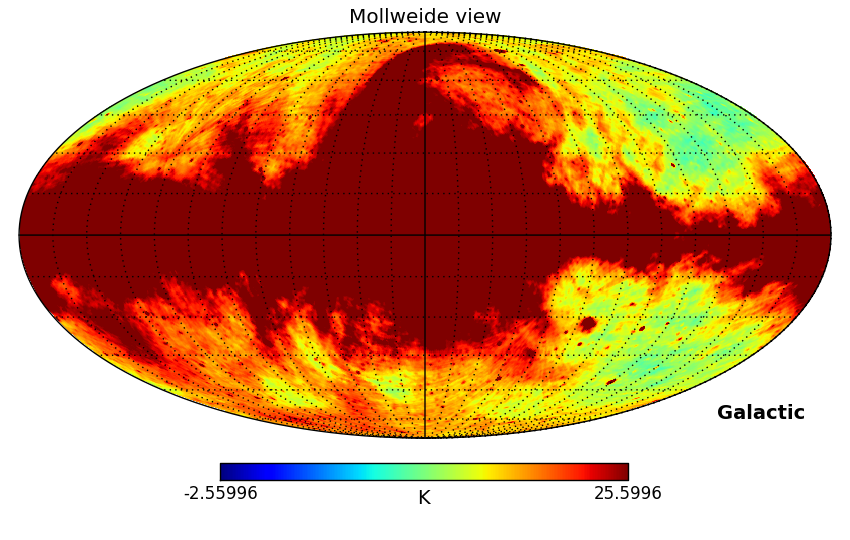}
\end{minipage}
\begin{minipage}[b]{0.33\linewidth}
\includegraphics[width=\linewidth]{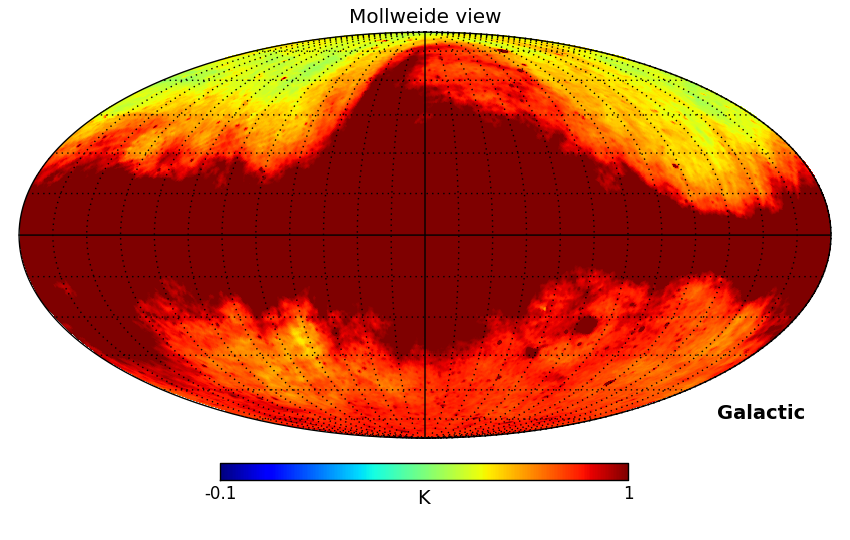}
\end{minipage}
\begin{minipage}[b]{0.33\linewidth}
\includegraphics[width=\linewidth]{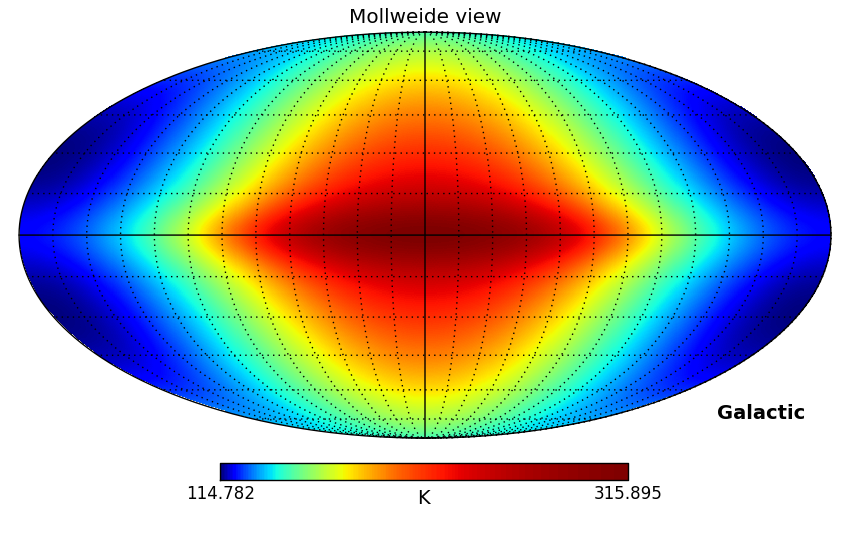}
\end{minipage}
\begin{minipage}[b]{0.33\linewidth}
\includegraphics[width=\linewidth]{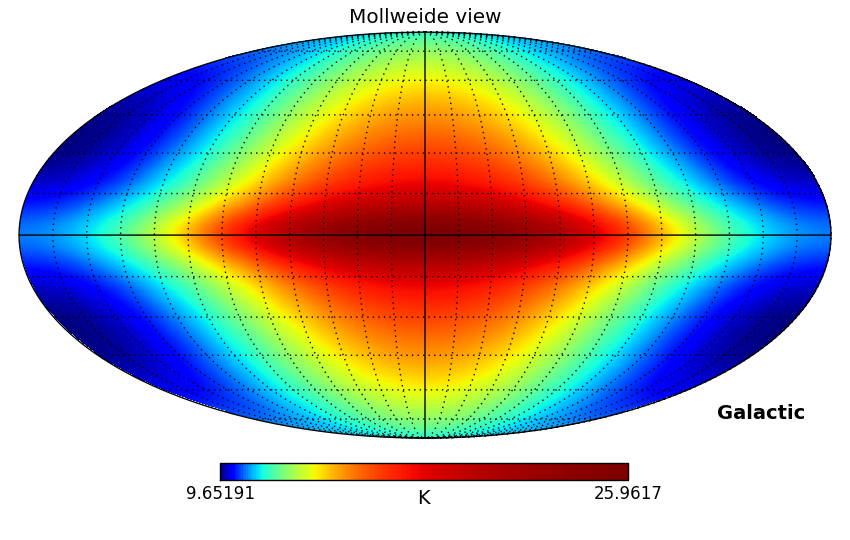}
\end{minipage}
\begin{minipage}[b]{0.33\linewidth}
\includegraphics[width=\linewidth]{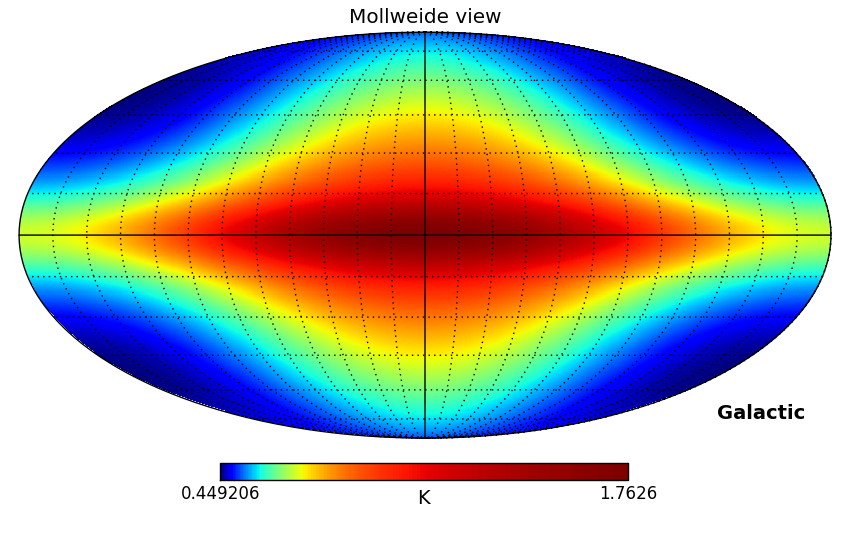}
\end{minipage}
\begin{minipage}[b]{0.33\linewidth}
\includegraphics[width=\linewidth]{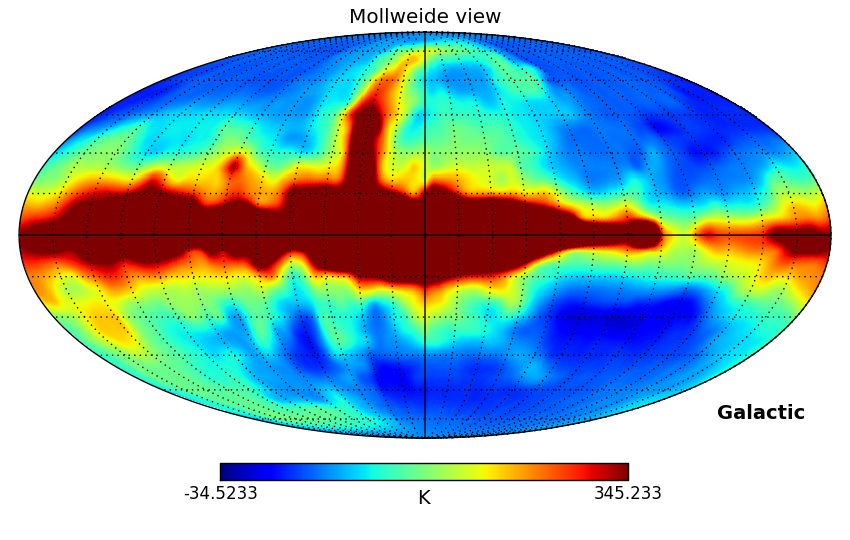}
\end{minipage}
\begin{minipage}[b]{0.33\linewidth}
\includegraphics[width=\linewidth]{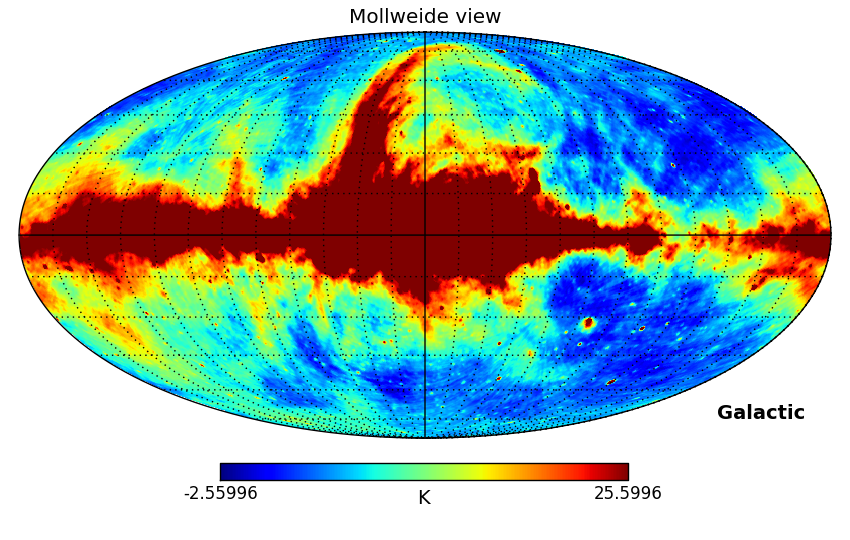}
\end{minipage}
\begin{minipage}[b]{0.33\linewidth}
\includegraphics[width=\linewidth]{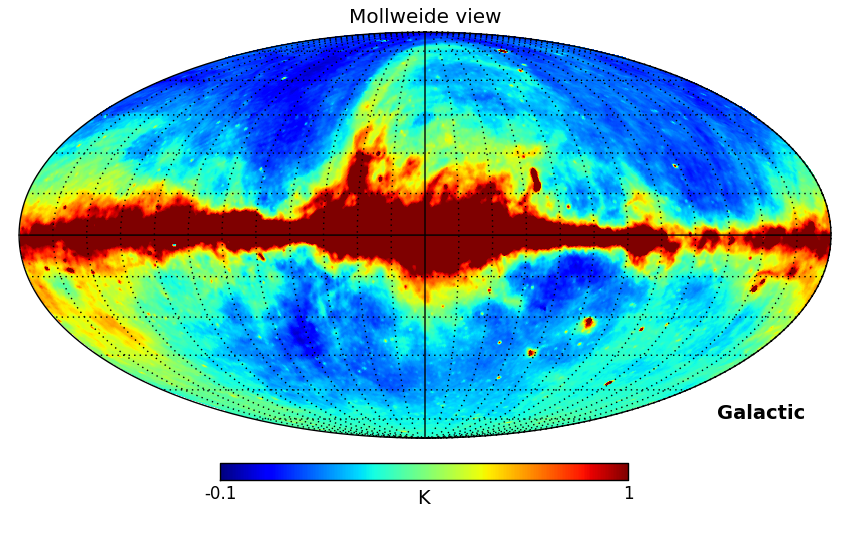}
\end{minipage}
\caption{Results of the model fit to the cosmic radio background at 150, 408 and 1420~MHz.  The column of three panels on the left are for 150~MHz, the middle column corresponds to 408 MHz and the column on the right is for 1420~MHz. In the three upper panels are shown the all-sky maps, in the middle panels we show the models and in the lower panels are shown the remainders on subtraction of the model from the data.  The top and bottom panels are in the same linear scale for ease of comparison, the middle panel is a histogram equalized color representation to show the apparent structure in the model clearly.}       
\label{fig:fit_results}
\end{figure}
\end{landscape}

\begin{figure}
\begin{minipage}[Ht]{0.5\linewidth}
\includegraphics[angle=-90, width=\linewidth]{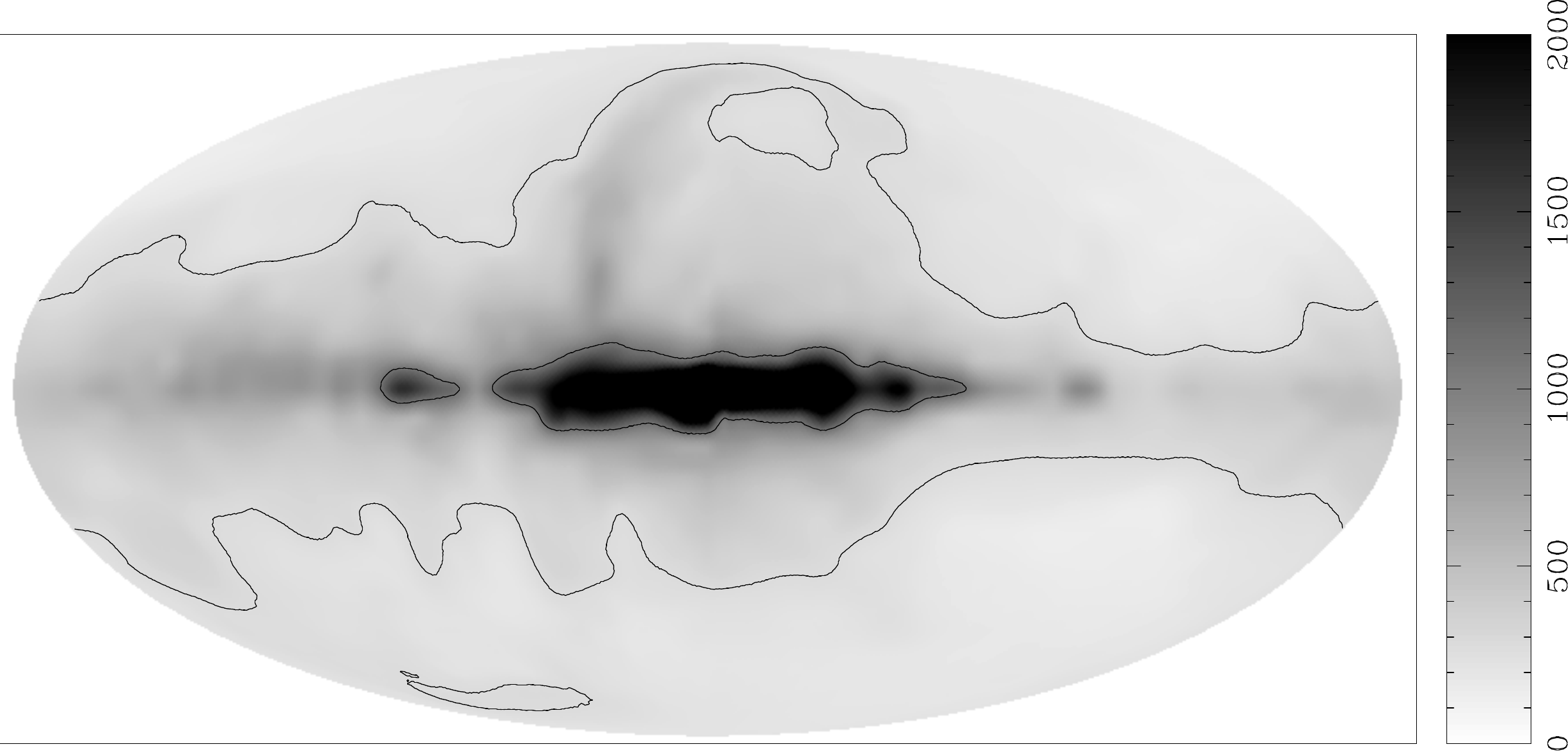}
\end{minipage}
\begin{minipage}[Ht]{0.5\linewidth}
\includegraphics[angle=-90, width=\linewidth]{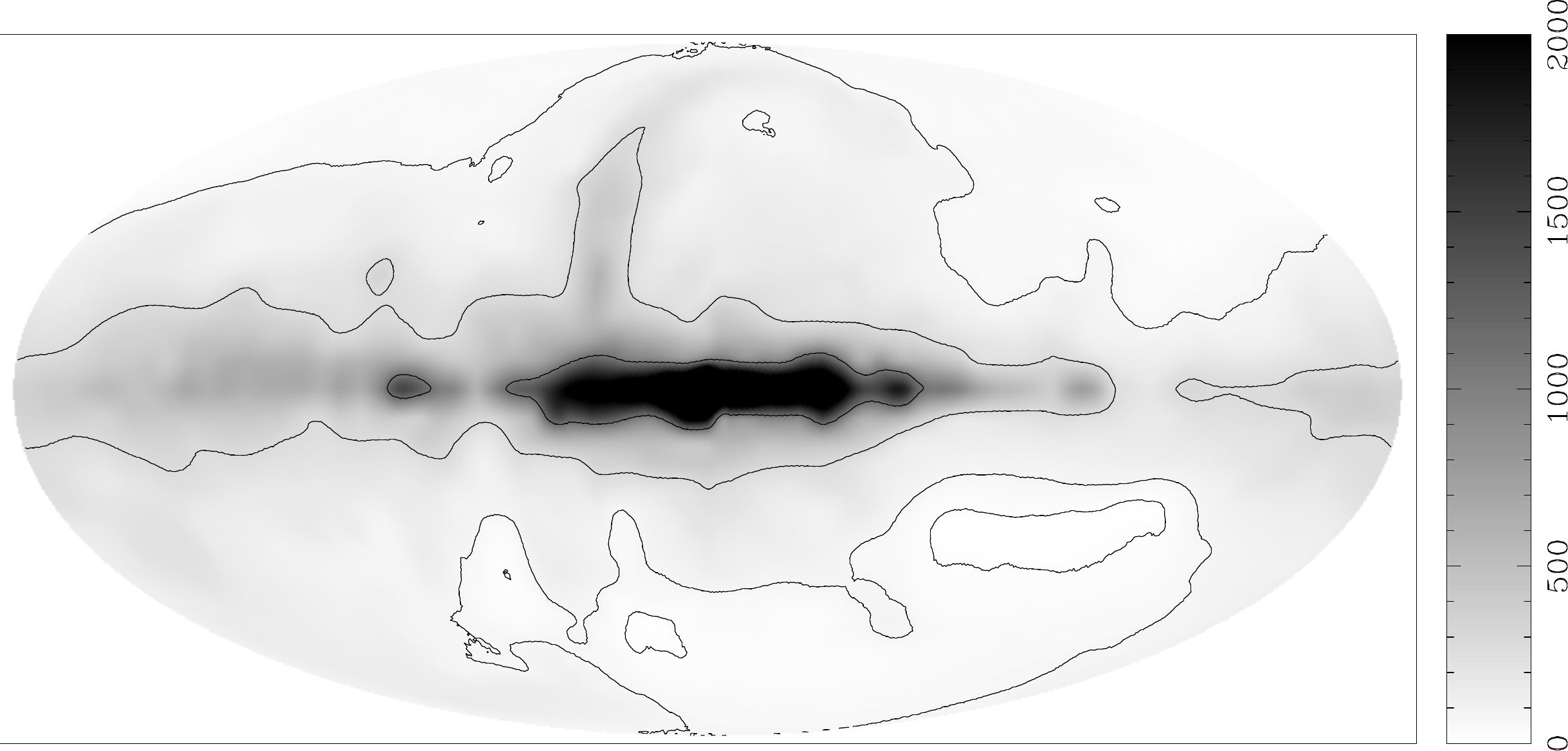}
\end{minipage}
\begin{minipage}[Ht]{0.5\linewidth}
\includegraphics[angle=-90, width=\linewidth]{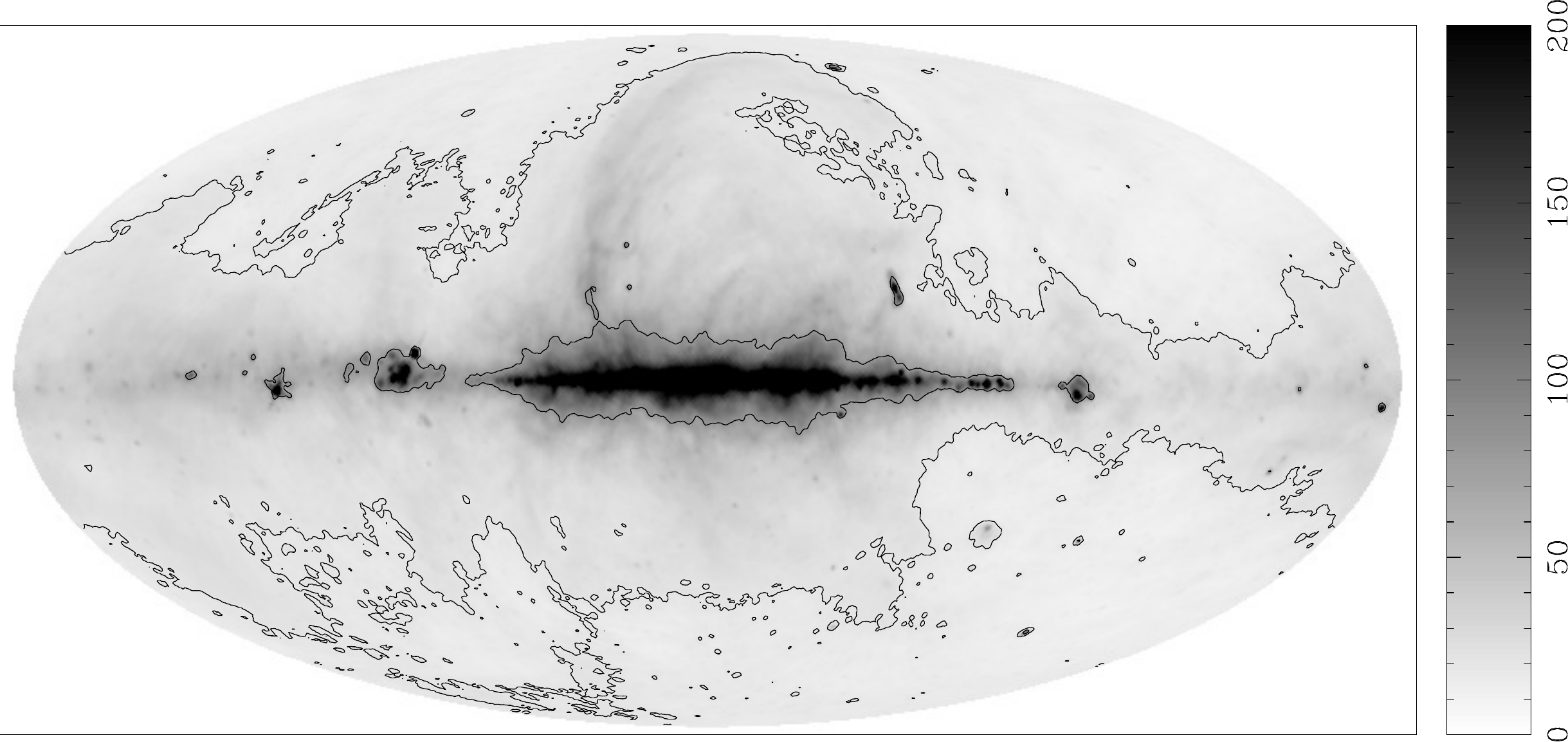}
\end{minipage}
\begin{minipage}[Ht]{0.5\linewidth}
\includegraphics[angle=-90, width=\linewidth]{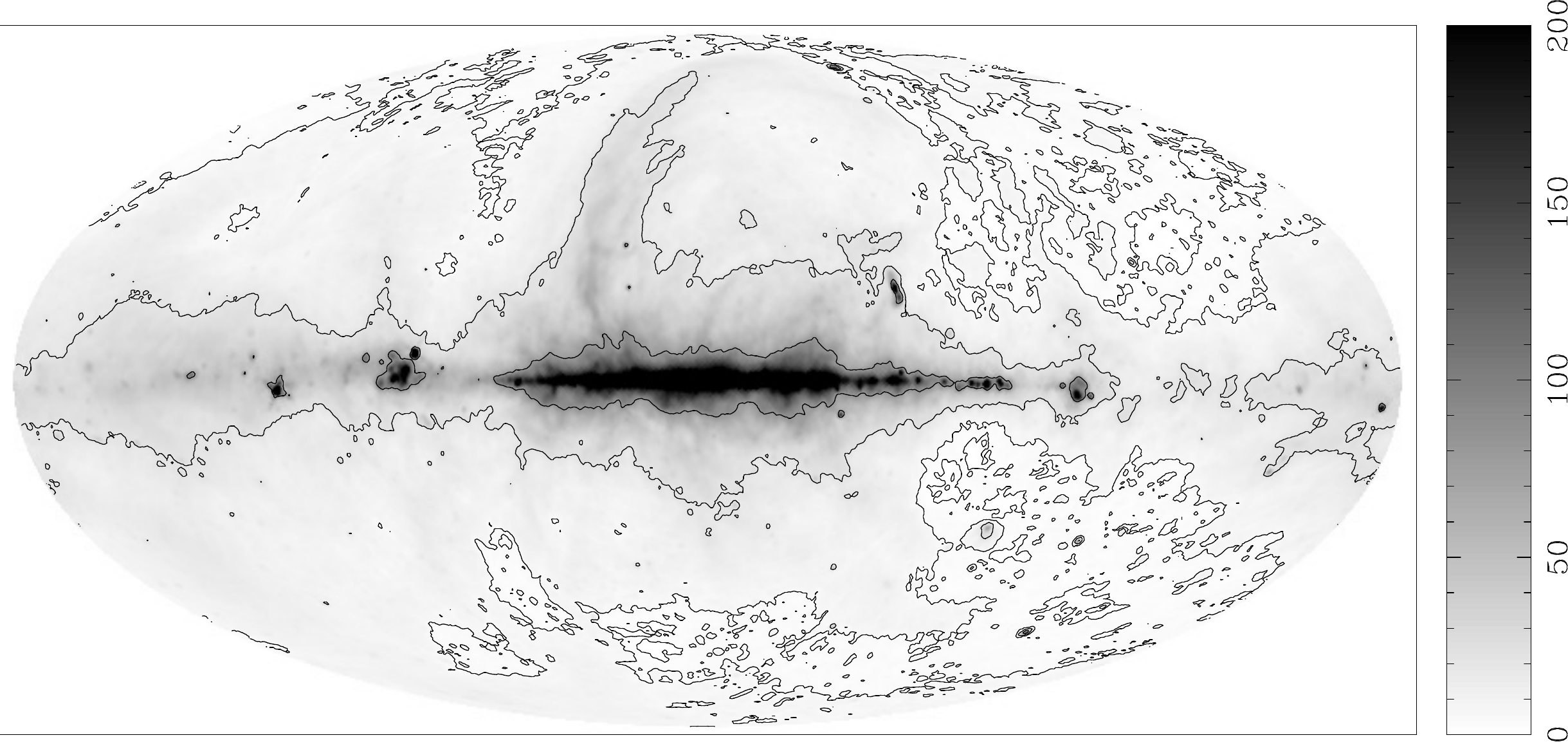}
\end{minipage}
\begin{minipage}[Ht]{0.5\linewidth}
\includegraphics[angle=-90, width=\linewidth]{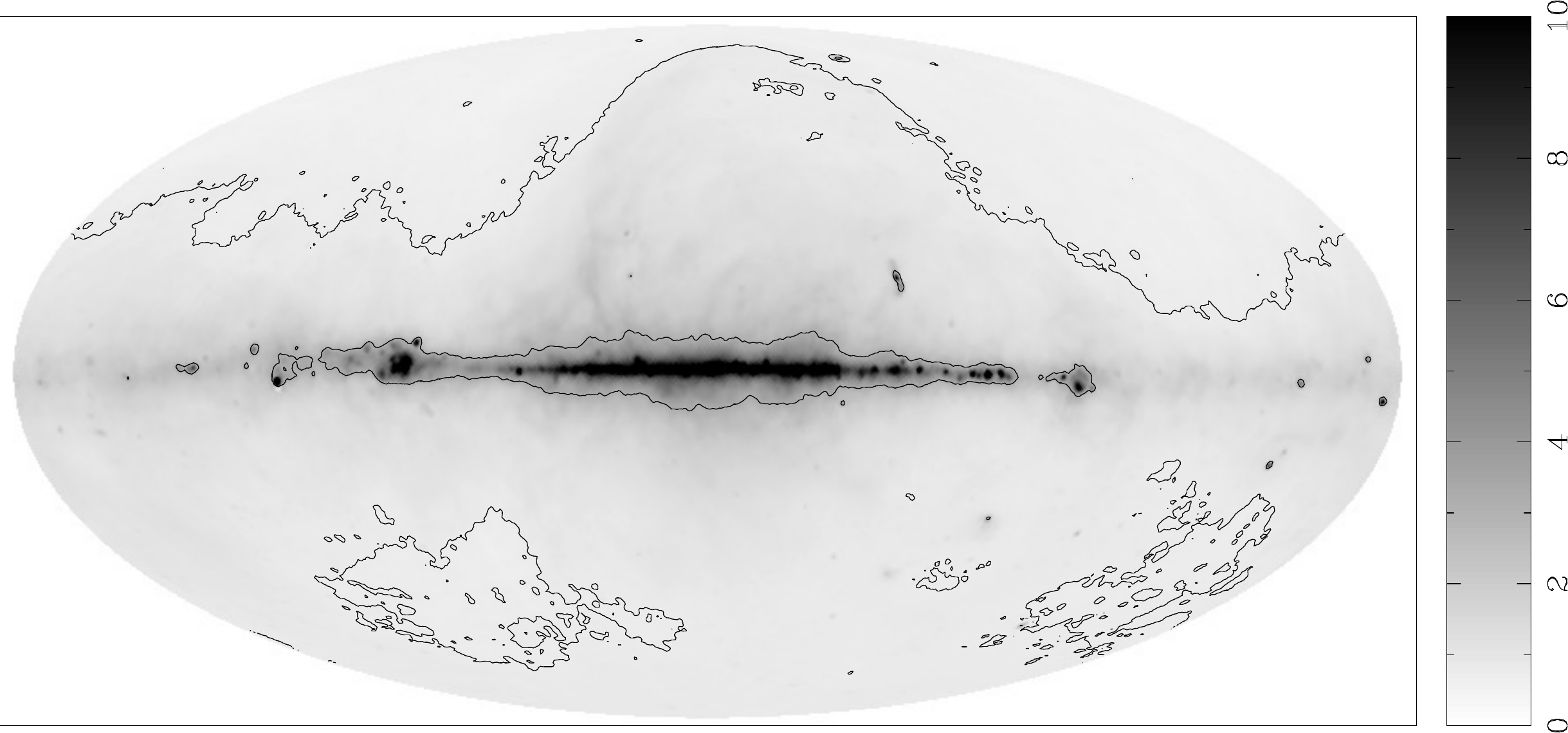}
\end{minipage}
\begin{minipage}[Ht]{0.5\linewidth}
\includegraphics[angle=-90, width=\linewidth]{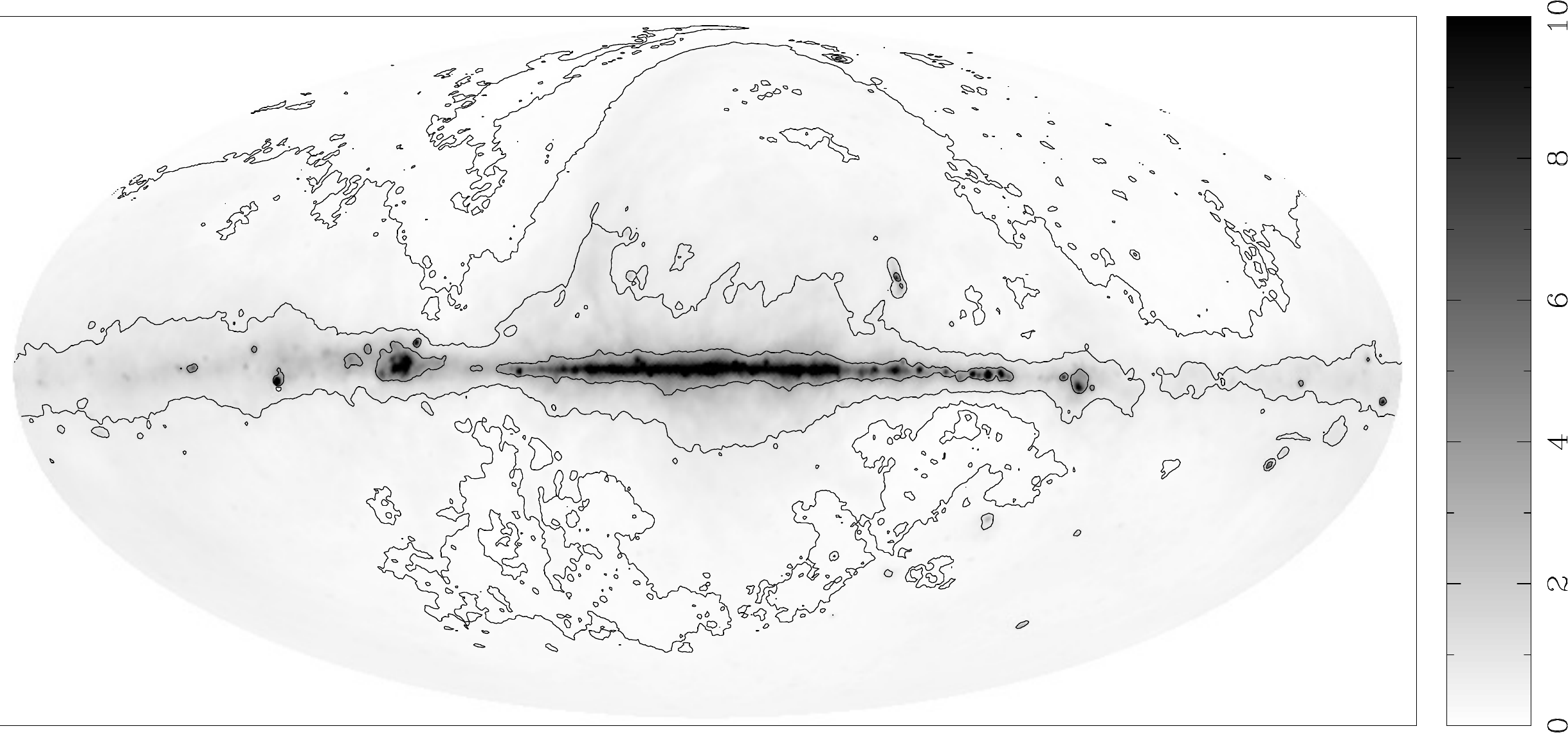}
\end{minipage}
\caption{Contour display of the all-sky images and remainders.  The all-sky images are shown in the left column and corresponding remainder images in the column on the right.  The top row is for 150~MHz, the middle row is 408~MHz and the row of two images at the bottom are at 1420~MHz.  Contours are at 1, 4, 16 and 64 times the contour unit, which is twice the final threshold $\tau$ and has values 10, 1.28 and 0.05~K respectively at 150, 408 and 1420~MHz.  Grey scale images span the range indicated by the wedge associated with each image, grey scale is in Kelvin brightness temperature.}
\label{fig:fit_results2}
\end{figure}

The results of model fits at 150, 408 and 1420~MHz are given in Table~\ref{fit_parameters}.  For comparison, the slope $a_{1}$ and intercept $a_{0}$ from fits of plan parallel slab models to each of the all-sky maps separately is also given in the top two rows of the Table.  In Fig.~\ref{fig:fit_results} we show the all-sky images---with CMB monopole of 2.725~K subtracted throughout---in the upper panels, the best-fit models in the middle panels and the remainders in the lower panels.   For quantitative display of the quality of the fits, contour plots of the all-sky images and remainders are shown separately in Fig.~\ref{fig:fit_results2}. As compared to a simple slab model, the more realistic model and fitting method we have adopted suggests substantially lower values for the extragalactic uniform brightness.  

\begin{table}[!htbp]
\caption{Parameters that describe the decomposition of the all-sky maps.}
\label{fit_parameters}
\centering
\begin{tabular}{lrrr}
\tableline\tableline
 & 150 MHz & 408 MHz & 1420 MHz \\
\tableline
Slope $a_1$ for a slab model & 69~K & 5.0~K & 0.17~K \\
Intercept $a_0$ & 143~K & 13~K & 0.62~K \\
& & & \\ 
Semi-major axis\tablenotemark{a} of the spheroid & 1.60 & 1.56 & 2.1 \\
Semi-minor axis\tablenotemark{a} of the spheroid & 0.29 & 0.24 & 0.37 \\
Radius\tablenotemark{a} of the sphere & 2.39 & 2.14 & 1.8 \\
Axial ratio of the spheroid & 5.6 & 6.4 & 5.6 \\
 & & & \\
Brightness of spheroid in the plane\tablenotemark{b} & 69 K & 7 K & 0.79 K \\
Brightness of spheroid towards the poles\tablenotemark{b} & 12 K & 1.1 K & 0.14 K \\
Brightness of the sphere\tablenotemark{b} & 129 K & 6.9 K & 0.30 K \\
& & & \\
Background brightness from the optimization & 21 K & 4.5 K & 0.14 K \\
& & & \\
Mean of the Markov chain sampling of the background brightness & 28 K & 2.5 K & 0.12 K \\
\tableline
\end{tabular}
\tablenotetext{a}{In units of the radius of the solar circle.}
\tablenotetext{b}{For an observer at the Galactic center.}
\end{table}

The fits gave mean values of 1.75 $\omega_\odot$ and 0.3 $\omega_\odot$ for the semi-major and semi-minor axes of the spheroid, and a mean value of 2.1 $\omega_\odot$ for the radius of the spherical halo.   The axial ratio of the spheroidal components, which describe the thick disc, is about 6 at all frequencies.  At 1420~MHz and for an observer located at the Galactic center, the halo has a brightness of 0.3~K  toward the pole and the disc has a brightness of 0.14~K.  The halo component appears to progressively dominate the total Galactic brightness at lower frequencies: at 150~MHz the halo has a brightness 129~K towards the pole and the disc has a brightness of 12~K.  This trend is consistent with previous analyses that indicated a steeper spectral index for the halo compared to the disc \citep{Web75}.

\subsection{An MCMC analysis of the distribution in the model parameters}

We have used the affine-invariant ensemble sampler for Markov Chain Monte Carlo (MCMC) proposed by \citet{Goo10} and implemented in Python by \citet{For13} to derive the distribution in the model parameters consistent with the constraints defined above.  The natural logarithm of  the posterior probability corresponding to any choice of parameters was defined to be proportional to negative of the statistic defined above.  For the six parameter model, we adopted 20 `walkers' representing the computation of 20 coupled Markov chains.   The initial vector positions of the `walkers' in the parameter space was adopted to be a fairly tight ball randomly positioned around the location of the solution from the optimization; nevertheless, the `walkers' did spread out and sample the wide parameter space allowed by the priors.  Initial chain lengths corresponding to about 15 correlations lengths were dropped as a `burn-in' phase.  The acceptance fraction was found to be reasonable, about 0.3.  

\begin{figure}[!Ht]
\begin{minipage}[!Ht]{0.33\linewidth}
\includegraphics[width=\linewidth]{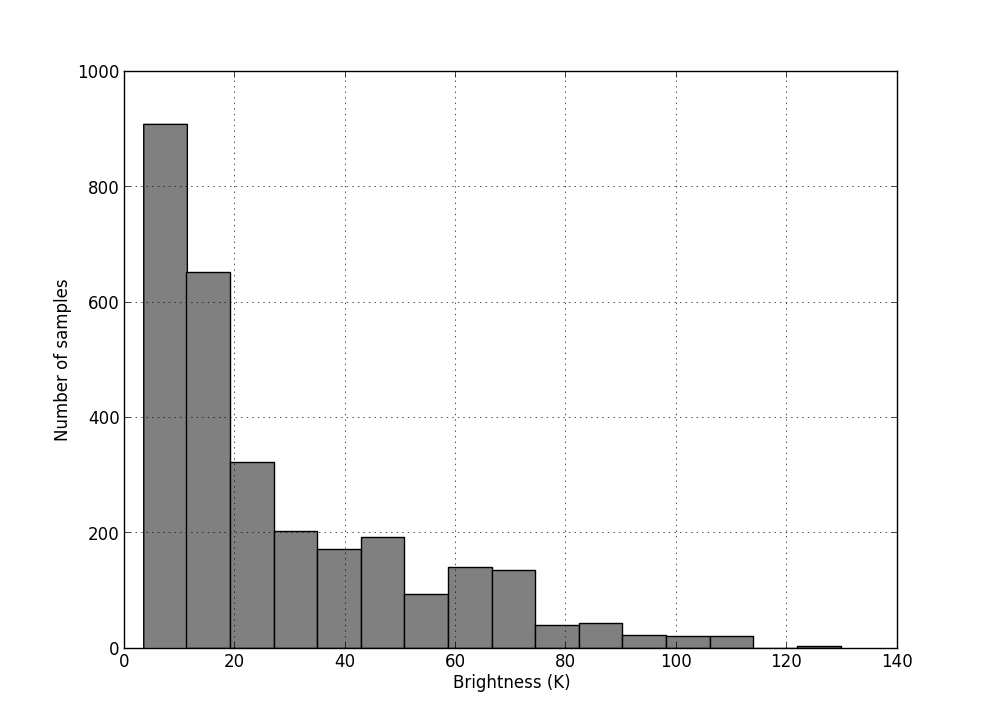}
\end{minipage}
\begin{minipage}[!Ht]{0.33\linewidth}
\includegraphics[width=\linewidth]{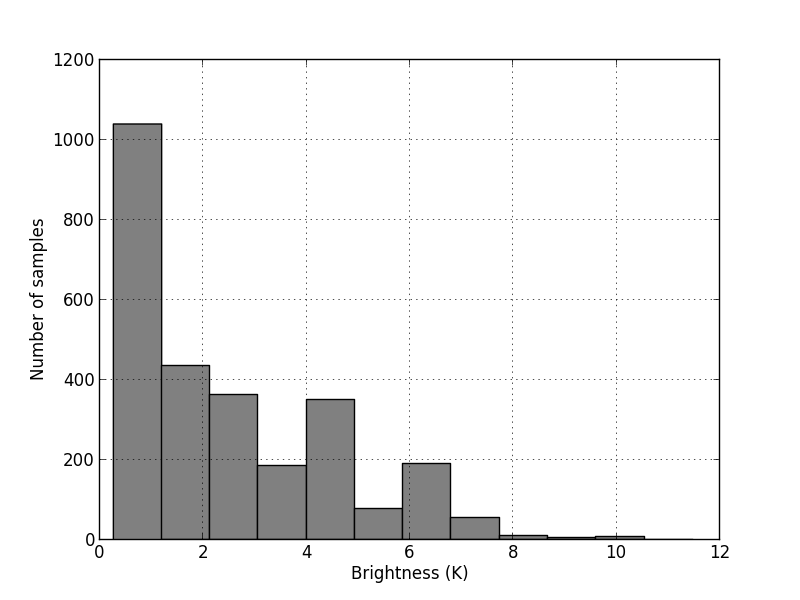}
\end{minipage}
\begin{minipage}[!Ht]{0.33\linewidth}
\includegraphics[width=\linewidth]{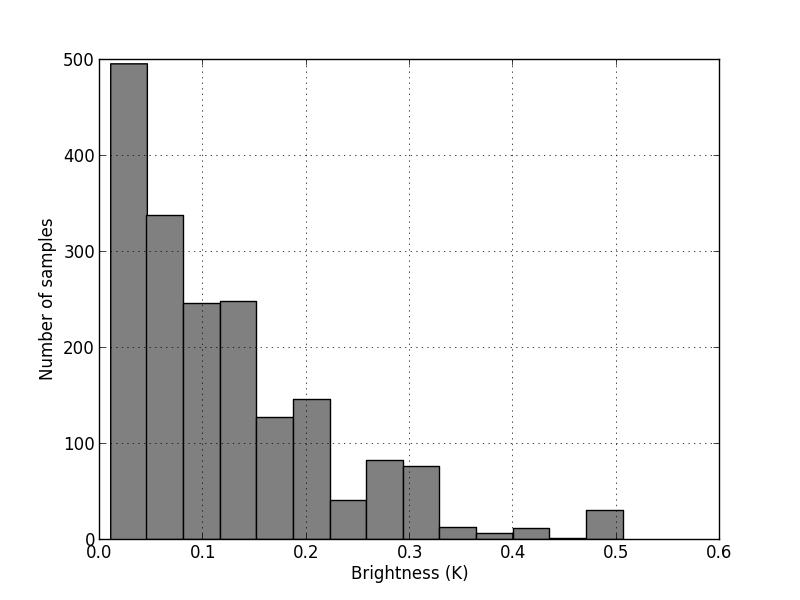}
\end{minipage}
\caption{Histograms of the distribution in brightness of the uniform extragalactic background from the MCMC chains.  The sampling is consistent with the priors and the adopted statistic for the posterior probability.  The three histograms are for the brightness at three frequencies: the panel on the left is for 150~MHz, in the center is 408~MHz and on the right is the histogram for 1420~MHz. }       
\label{fig:hist_tbg}
\end{figure}

\begin{figure}[!Hb]
\begin{minipage}[!Hb]{0.33\linewidth}
\includegraphics[width=\linewidth]{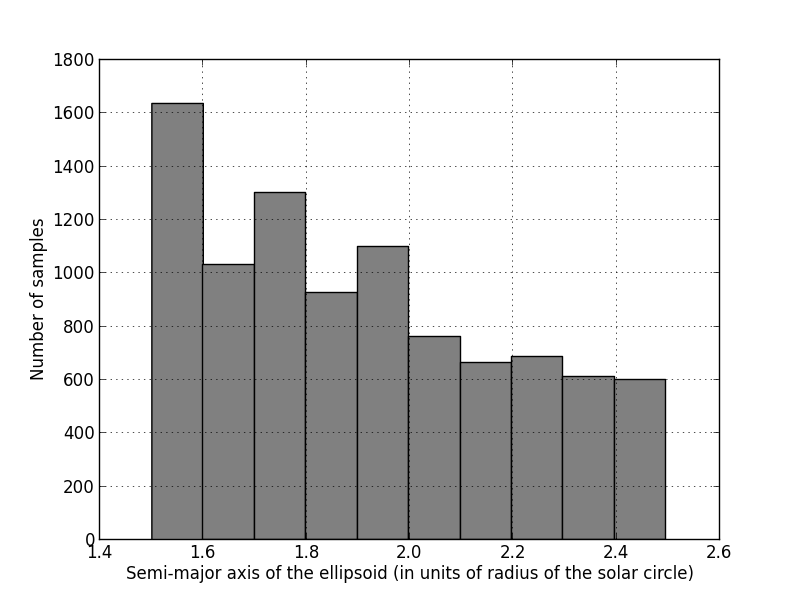}
\end{minipage}
\begin{minipage}[!Hb]{0.33\linewidth}
\includegraphics[width=\linewidth]{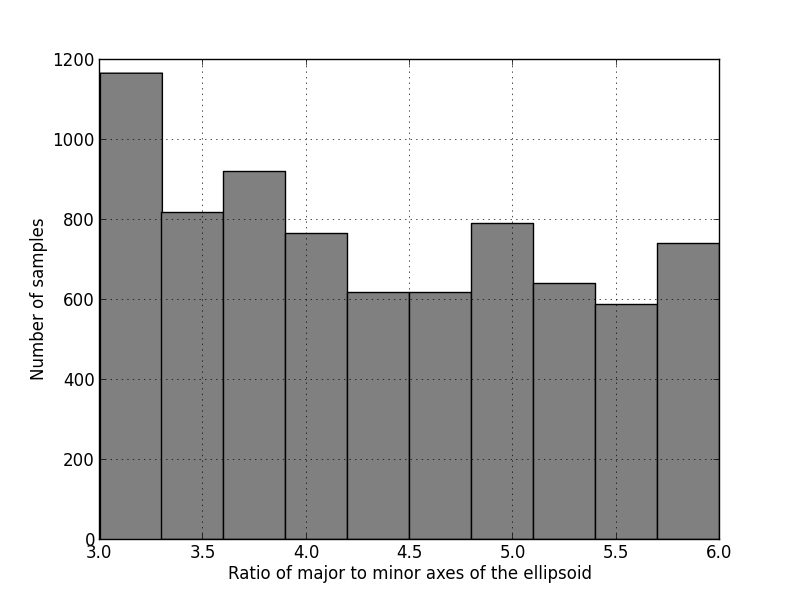}
\end{minipage}
\begin{minipage}[!Hb]{0.33\linewidth}
\includegraphics[width=\linewidth]{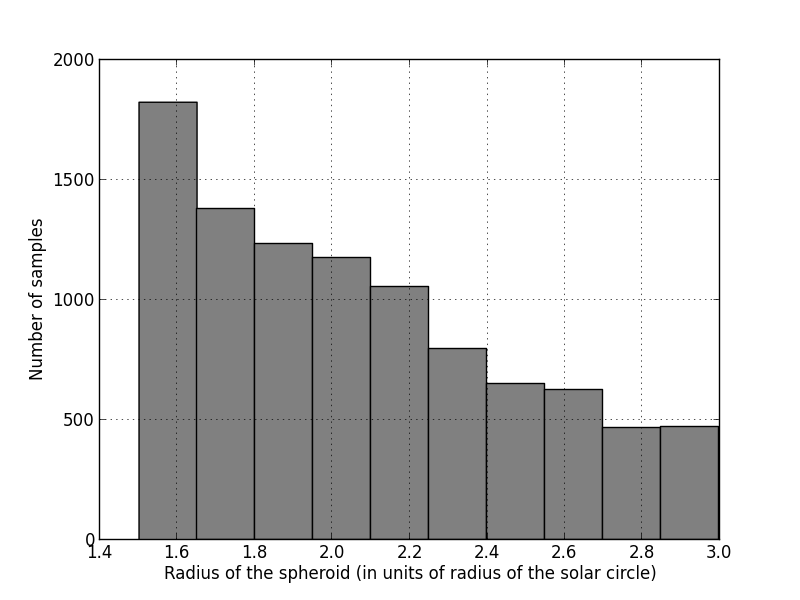}
\end{minipage}
\caption{Histograms of the distribution in parameters describing the model for the Galactic emission.  In the left panel is the histogram of the distribution in the semi-major axis of the spheroidal component and in the panel on the right is the histogram of the distribution of the radius of the spherical halo component; both these are in units of the radius of the solar circle.  In the panel at the center is the histogram of the ratio of semi-major to semi-minor axes of the spheroidal component.}       
\label{fig:hist_model}
\end{figure}

We show in Fig.~\ref{fig:hist_tbg} histograms of the distribution of the Markov chain values of the uniform brightness.  The mean values of the Markov chains suggest values for the brightness of the background of 28, 2.5 and 0.12~K respectively at 150, 408 and 1420~MHz.  The adopted priors and constraints that are embodied in the posterior probability function appears to prefer low values for the uniform brightness.  Once again, as anticipated, these mean values are substantially below the values corresponding to the intercepts $a_{0}$ listed in the second row of Table~\ref{fit_parameters}.  

We show in Fig.~\ref{fig:hist_model} histograms of the distribution of the Markov chain values for the major axis of the disc, ratio of major to minor axes of the disc and the radius of the spherical halo.  All of these distributions sample the entire range allowed by the priors: the posterior likelihood based on the statistic we have adopted does not strongly constrain the sampling to a relatively small region of the allowed parameter space.  Although the priors are uniform, the sampling distributions prefer smaller values for the semi-major axis of the disc and relatively smaller ratios for the semi-major to semi-minor axes: i.e. thick discs are preferred to thin discs.  The mean value of the sampling distribution of the semi-major axis is 1.9 times the radius of the solar circle and the mean ratio of major to minor axis is 4.4.   The distribution for the radius of the spherical halo is skewed towards smaller radii and this distribution has a mean radius of 2.1 times the radius of the solar circle.

Clearly the adopted posterior probability function, which is based on the function that is minimized for the optimization, allows for a wide range of plausible values for the model parameters.  Nevertheless, the extragalactic background brightness is fairly well constrained toward small values, which are substantially smaller than that indicated if a plane parallel slab model were adopted for the Galaxy.  The mean values of the Markov chain sampling distributions are also listed in Table~\ref{fit_parameters}.

\section{Discussion and Summary}

\citet{Ger08} estimated the contribution from discrete sources to the extragalactic background at 150~MHz to be 39~K, at 408~MHz to be 2.65~K and at 1420~MHz to be 0.094~K all based on a model for the source counts that included multiple populations and a frequency dependence.  Separately, motivated by the suggestion that there might be an excess uniform extragalactic background, \citet{Ver11} examined the range allowed for the extragalactic uniform brightness constrained by observed source counts at different frequencies, and found the range to be 18.1--29.4~K at 150~MHz, in the range 1.85--3.0~K at 408~MHz and in the range 0.11--0.18~K at 1420~MHz.  The Markov chain sampling distribution of the background brightness has a mean (the last row of Table~\ref{fit_parameters}) that is within the range of extragalactic brightness estimated from the source counts.

Fitting the all-sky maps to a model with a single component in the form of a plane parallel slab yields intercepts of 143, 12.9 and 0.62~K respectively at 150, 408 and 1420~MHz, after excluding the CMB monopole contribution.  However, the more comprehensive procedure for resolving the maps into various components with astrophysically motivated models we adopt here yield estimates of 21--28, 2.5--4.5 and 0.12--0.14~K for the extragalactic uniform brightness, which are smaller by factors of 3--7 and are consistent with the brightness estimated from the observed source counts.  Clearly, realistic modeling of the Galactic emission and an effective procedure for decomposition of the sky maps are necessary in order to answer the question of whether or not there is an unaccounted `excess' extragalactic brightness in the cosmic radio background.  In this context we would like to emphasize the need for using images with substantial sky coverage for the decomposition, since any Galactic halo is a large scale feature on the sky dominated by a dipole anisotropy.

The absolute values we have derived for the brightness of the different components, including that of the uniform extragalactic sky, will suffer uncertainties owing to the errors in the calibration of the all-sky maps we have used. The 150-MHz image has uncertainties of $\pm40$~K  in the zero point and 5--7$\%$ in absolute scale.  The 408-MHz image has a quoted accuracy of $\pm3$~K  in the zero point and $10\%$ in absolute scale; the $3 \sigma$ rms noise is 50~mK. The 1420~MHz map has quoted accuracy of $\pm 0.5$~K in zero point and 5\% in absolute scale; the $3 \sigma$ rms noise is 50~mK.  The rms noise is relatively small; the dominant errors in the derivation of the absolute brightness of the different components is the limited accuracy in the zero points of the individual maps.  Unfortunately, the quoted errors in zero point are substantial and comparable to or greater than the derived brightness of the extragalactic uniform brightness, which implies that it is really pointless to fit to these maps with an aim of accurately determining the value of the uniform extragalactic brightness unless their calibration is revisited and improved significantly.  This is indeed an excellent motivation for improved absolute measurements of the cosmic radio background at multiple frequencies at least to establish the scale and zero-point of these low frequency all-sky images so that the uniform extragalactic background may be accurately known.

TRIS radiometers measured the absolute temperature of an annular patch of the sky at 0.6, 0.82 and 2.5~GHz \citep{Zan08}.  These were used in an attempt to improve the calibration of the \citet{Has82} 408-MHz map \citep{Tar08}.  TRIS measurements were limited in sky coverage and limiting their analysis to that sky region they estimated that the 408-MHz map required a correction of $+3.9 \pm 0.6$~K of the zero point.  Such a correction would enhance the estimated value of the uniform extragalactic background by 3.9~K, making the derived background brightness about a factor of two over the range allowed by the source counts.  A revision of the zero point would also revise upward the derived value of the intercept $a_0$ and the estimate for the extragalactic brightness based on the multi component modeling of the Galactic emission would be a factor of two below the estimate based on a plane parallel slab model.

The dipole and, to a lesser extent, higher order multipoles arising from any Galactic spherical halo would be a contaminant that needs to be modeled and subtracted while estimating the anisotropy in the CMB.  Our optimization suggests a brightness for the spherical halo that is fit by a relation $T_{\rm halo} = 0.72~\nu_{\rm GHz}^{-2.7}$~K, where $T_{\rm halo}$ is the brightness of the halo as viewed from the center of the Galaxy.  The dipole anisotropy would have a brightness temperature amplitude of about half $T_{\rm halo}$ for a halo that has a radius twice the radius of the solar circle.  If the temperature spectral index continued to be $-2.7$, the dipole amplitude would be about $20~\mu$K at 50~GHz observing frequency.  The synchrotron model for the Galactic emission must necessarily have to include the halo component, otherwise a frequency dependent dipole would be present in the foreground subtracted images made by WMAP and PLANCK.

Admittedly simulated annealing does not necessarily yield a global minimum for the function that is being optimized.  And once the minimization is trapped in a local minimum at the first step, subsequent stages of downhill simplex may not allow the solution to jump to a distant global minimum.  Therefore, the models we have derived might well be local minima and indicative of a plausible decomposition.  For this reason, the analysis presented herein may be interpreted as demonstrating that an `excess' uniform background is not an imperative of the data but depends on the model components chosen for the Galactic emission. Not withstanding these cautionary remarks, we may note that the careful choice of the priors and the similarity of the conclusions drawn by the optimization and Monte Carlo methods suggest that the results derived here are a good representation of the astrophysical situation.

We conclude that there is no compelling evidence for an unexplained `excess' uniform background: careful and realistic modeling of the Galactic emission, which include a spherical halo, a flattened spheroidal disc, and a set of highly anisotropic features attributable to sources, filaments, SNR's, etc., can account for all of the anisotropic Galactic emission leaving an isotropic component consistent with known extragalactic source counts.



\acknowledgments


\begin{thebibliography}{}
\bibitem[Bennett et al.(2003)]{Ben03} 
Bennett, C.~L., Hill, R.~S., Hinshaw, G., et al.\ 2003, \apjs, 148, 97
\bibitem[Beuermann et al.(1985)]{Beu85} 
Beuermann, K., Kanbach, G., \& Berkhuijsen, E.~M.\ 1985, \aap, 153, 17 
\bibitem[Biermann \& Harms (2013)]{Bie13}
Biermann, P.~L., \& Harms, B.~C.\ 2013, arXiv:1305.0498
\bibitem[Clark, Brown \& Alexander (1970)]{Cla70}
 Clark, T. A., Brown, L. W., \& Alexander, J.K. 1970, \nat, 228, 847
 \bibitem[Condon et al.(2012)]{Con12} 
 Condon, J.~J., Cotton, W.~D., Fomalont, E.~B., et al.  2012, \apj, 758, 23 
\bibitem[deOliveira-Costa et al. (2008)]{deO08}
de Oliveira-Costa, A., Tegmark, M., Gaensler, B. M., Jonas, J., Landecker, T. L., \& Reich, P. 2008, \mnras, 388, 247
\bibitem[Fixsen et al.(2011)]{Fix11} 
Fixsen, D.~J., Kogut, A., Levin, S., et al.\ 2011, \apj, 734, 5 
\bibitem[Foreman-Mackey et al.(2013)]{For13} 
Foreman-Mackey, D., Hogg, D.~W., Lang, D., \& Goodman, J.\ 2013, \pasp, 125, 306
\bibitem[Fornengo et al.(2011)]{For11} 
Fornengo, N., Lineros, R., Regis, M., \& Taoso, M.\ 2011, Physical Review Letters, 107, 261302 
\bibitem[Gervasi et al.(2008)]{Ger08} 
Gervasi, M., Tartari, A., Zannoni, M., Boella, G., \& Sironi, G.\ 2008, \apj, 682, 223 
\bibitem[Goodman \& Weare (2010)]{Goo10}
Goodman, J., \& Weare, J.\ 2010, Comm. App. Math. Comp. Sci., 5, 65
\bibitem[G{\'o}rski et al.(2005)]{Gor05} 
G{\'o}rski, K.~M., Hivon, E., Banday, A.~J., et al.\ 2005, \apj, 622, 759
\bibitem[Haslam et al.(1982)]{Has82} 
Haslam, C.~G.~T., Salter, C.~J., Stoffel, H., \& Wilson, W.~E.\ 1982, \aaps, 47, 1
\bibitem[Holder (2012)]{Hol12}
Holder, G. 2012, \apj\ submitted, arXiv:1207.0856
\bibitem[Jansky (1933)]{Jan33}
Jansky, K. G. 1933, \nat, 132, 66
\bibitem[Kirkpatrick et al.(1983)]{Kir83} 
Kirkpatrick, S., Gelatt, C.~D., \& Vecchi, M.~P.\ 1983, Science, 220, 671
\bibitem[Kogut et al.(2011)]{Kog11} 
Kogut, A., Fixsen, D.~J., Levin, S.~M., et al.\ 2011, \apj, 734, 4
\bibitem[Landecker \& Wielebinski(1970)]{Lan70}
Landecker, T.~L., \& Wielebinski, R.\ 1970, Australian Journal of Physics Astrophysical Supplement, 16, 1 
\bibitem[Nelder \& Mead (1965)]{Nel65}
Nelder, J.~A., Mead, R.\  1965,  Computer Journal, 7, 308
\bibitem[Platania et al.(2003)]{Pla03} 
Platania, P., Burigana, C., Maino, D., et al.\ 2003, \aap, 410, 847 
\bibitem[Reich(1982)]{Rei82} 
Reich, W.\ 1982, \aaps, 48, 219
\bibitem[Reich \& Reich(1986)]{Rei86} 
Reich, P., \& Reich, W.\ 1986, \aaps, 63, 205 
\bibitem[Reich et al.(2001)]{Rei01} 
Reich, P., Testori, J.~C., \& Reich, W.\ 2001, \aap, 376, 861
\bibitem[Salter \& Brown(1988)]{Sal88}
Salter, C.~J. \& Brown, R.~L.\ 1988, in \emph{Galactic and Extragalactic Radio Astronomy} (eds. G. L. Verschuur and K. I Kellerman with assistance of E. Bouton) 2nd Edition, Berlin: New York: Springer-Verlag
\bibitem[Seiffert et al.(2011)]{Sei11} 
Seiffert, M., Fixsen, D.~J., Kogut, A., et al.\ 2011, \apj, 734, 6
\bibitem[Singal et al.(2010)]{Sin10} Singal, J., Stawarz, 
{\L}., Lawrence, A., \& Petrosian, V.\ 2010, \mnras, 409, 1172 
\bibitem[Tartari et al.(2008)]{Tar08} 
Tartari, A., Zannoni, M., Gervasi, M., Boella, G., \& Sironi, G.\ 2008, \apj, 688, 32
\bibitem[Vernstrom et al.(2011)]{Ver11} 
Vernstrom, T., Scott, D., \& Wall, J.~V.\ 2011, \mnras, 415, 3641
\bibitem[Webster(1975)]{Web75} 
Webster, A.\ 1975, \mnras, 171, 243
\bibitem[Zannoni et al.(2008)]{Zan08} 
Zannoni, M., Tartari, A., Gervasi, M., et al.\ 2008, \apj, 688, 12
\end{thebibliography}
\end{document}